\documentstyle[emulateapj,epsf]{article}

\begin{document}

\title{The mass surface density in the local disk and the 
       chemical evolution of the Galaxy}

\author{Donatella Romano, Francesca Matteucci\altaffilmark{1}, Paolo Salucci}

\affil{SISSA/ISAS, via Beirut 2-4, I-34014 Trieste, Italy; 
       romano, salucci@sissa.it}

\and

\author{Cristina Chiappini}
        
\affil{Departamento de Astronomia, Observat\'orio Nacional/CNPq, 
       Caixa Postal 23002, Rio de Janeiro, RJ, Brazil; 
       chiappin@danw.on.br}

\altaffiltext{1}{Dipartimento di Astronomia, Universit\`a di Trieste,
                 via G.B. Tiepolo 11, I-34131 Trieste, Italy; 
		 matteucci@ts.astro.it}

\begin{abstract}
We have studied the effect of adopting different values of the total baryonic 
mass surface density in the local disk at the present time, $\Sigma(R_\odot, 
t_{Gal})$, in a model for the chemical evolution of the Galaxy. We have 
compared our model results with the G-dwarf metallicity distribution, the 
amounts of gas, stars, stellar remnants, infall rate and SN rate in the solar 
vicinity, and with the radial abundance gradients and gas distribution in the 
disk. This comparison strongly suggests that the value of $\Sigma(R_\odot, 
t_{Gal})$ which best fits the observational properties should lie in the range 
50\,--\,75 $M_\odot$ pc$^{-2}$, and that values of the total disk mass surface 
density outside this range should be ruled out.
\end{abstract}


\section{INTRODUCTION}

Despite a number of estimates, obtained by means of several different methods, 
the value of the local column density of the Galactic disk in stars and gas, 
$\Sigma_{b}(R_\odot \simeq$ 8 kpc) = $\Sigma_{stars} + \Sigma_{gas}$, has been 
until recently very uncertain. Direct HST star counts imply a stellar column 
density of $\simeq 27$ $M_\odot$ pc$^{-2}$ (Flynn, Gould, \& Bahcall 1999), 
while the (H {\small I} + H$_2$) column density amounts to 7\,--\,13 $M_\odot$ 
pc$^{-2}$ (Dickey 1993; Flynn et al. 1999). Therefore, the surface density of 
the directly seen baryonic matter amounts to 35\,--\,40 $M_\odot$ pc$^{-2}$. 
However, the actual column density of the ``luminous'' disk could be higher 
because the above detection does not account for dead stars and brown dwarfs.
\par
The total disk surface density $\Sigma_{z}(R_\odot)$, including also any 
eventual disk dark matter component, can be determined from the motion 
of stars in the solar neighborhood perpendicular to the Galactic plane 
(e.g. Binney \& Tremaine 1987; Gilmore, Wyse, \& Kuijken 1989). From the 
kinematics of a sample of K-giants Bahcall, Flynn, \& Gould (1992) derived 
$\Sigma_{z}(R_\odot) = 84^{+30}_{-25}$ $M_\odot$ pc$^{-2}$ (at 1 $\sigma$ 
level). Notice that such a measure is a complex one and it is strongly 
dependent on the underlying assumptions. By employing different samples 
and/or analysis, lower values for $\Sigma_{z}(R_\odot)$ have been claimed: 
$\Sigma_{z}(R_\odot) = 46 \pm 9$ $M_\odot$ pc$^{-2}$ (Kuijken \& Gilmore 
1989) --- revised to $48 \pm 9$ $M_\odot$ pc$^{-2}$ (Kuijken \& Gilmore 1991), 
$\Sigma_{z}(R_\odot) = 52 \pm 13$ $M_\odot$ pc$^{-2}$ (Flynn \& Fuchs 1994). 
The local column density of stars and gas could be smaller, because these 
dynamical estimates may include some non-baryonic matter. Notice however that 
it is quite unlikely that there is non-baryonic matter in the disk, because 
non-baryonic dark matter is probably non dissipational. Any non-baryonic 
matter present in the disk is likely to be part of the dark halo, and this 
component is always subtracted from the dynamical estimates of the local 
column density.
\par
Fitting the Galaxy rotation curve (RC) with a mass model provides a further 
way of estimating the local column density of disk matter $\Sigma_{V}$. In 
fact, such a fit indicates what is the fraction $\beta$\footnote{It should 
be noted that the value for $\beta$ depends on the assumed profiles for the 
disk and the halo.} of the circular velocity due to the disk component (at 
the Sun's position). Then, $\beta\,V^2_\odot$ is compared with the circular 
velocity of an exponential thin disk, 
$V^2_{disk} = G^{-1}\,\Sigma_{V}\,R_{D}\,f(R_\odot/R_{D})$, 
with $f$ a known function of the Galactocentric distance of the Sun expressed 
in terms of disk length scales (see Freeman 1970), to yield $\Sigma_{V}$, 
providing that $R_\odot$ and $R_{D}$ are known. The results are far from 
unique, and reflect the presence of uncertainties in the mass modelling, in 
the actual shape of the Galactic RC, and in the basic structural parameters. 
The values found range between $\Sigma_{V} \sim 50$ $M_\odot$ pc$^{-2}$ and 
$\sim 100$ $M_\odot$ pc$^{-2}$ (Olling \& Merrifield 1998; Dehnen \& Binney 
1998), and moreover include any dark matter distributed like the stellar disk. 
In external galaxies of luminosity similar to the Galaxy $(L_{B} \sim 10^{11}$ 
$L_\odot)$, values of $\Sigma_{V}$ as high as $\simeq 90$ $M_\odot$ pc$^{-2}$ 
at $R$ = 8 kpc cannot be ruled out (Persic, Salucci, \& Stel 1996).
\par
Finally, let us note that if the dark matter halo in the Galaxy follows the 
standard cold dark matter (CDM) universal density profile (Navarro, Frenk, \& 
White 1997), and therefore at the Sun position $\rho_{SCDM} \propto R^{-2}$, 
then the local disk column density is required to be towards the low end of 
the above-cited values (e.g. Courteau \& Rix 1999). On the other hand, 
low-$\Omega$ CDM density profiles are not inconsistent with a baryon dominated 
region inside $R_\odot$, and then with much higher values for the local disk 
column density (Navarro 1998).
\par
Summarizing: all the above observations, often entangled with heavy 
theoretical assumptions, poorly constrain the value of the column density 
of the baryonic matter in the local disk, that may lie between 35 and 100 
$M_\odot$ pc$^{-2}$.
\par
However, very recently the results of the European Space Agency's Hipparcos 
mission have allowed a drastical reduction of this permitted range, by making 
available direct distances for the tracer stars. In particular, Cr\'ez\'e et 
al.\,(1998) and Holmberg \& Flynn (1998) have analyzed the A and F stars in 
the Hipparcos data set (some 10,000 stars) and derived greatly improved 
estimates of the total gravitating mass: the gravitating disk mass seems to 
be now firmly established at 50 to 60 $M_\odot$ pc$^{-2}$.
\par
In this paper we investigate how the gravitating mass of the disk influences 
the Galactic chemical evolution. Successful models for the chemical evolution 
of the Galaxy, in fact, require a star formation law depending on the total 
mass surface density (Tosi 1988; Matteucci \& Fran\c cois 1989; Chiappini, 
Matteucci, \& Gratton 1997), and the star formation rate (SFR) is a 
fundamental parameter in the evolution of galaxies.

\section{THE CHEMICAL EVOLUTION MODEL}

We adopt the two-infall chemical evolution model of Chiappini et al.\,(1997), 
to which we refer for an exhaustive explanation of the main assumptions and 
basic equations. Here we just review the features of the model most directly 
related to the aspects discussed in the present study.
\par
Briefly, the overall evolutionary scenario is the following: it is assumed 
that the Galaxy formed out of two main infall episodes; during the first 
episode, the primordial gas collapsed very quickly and formed the spheroidal 
components (halo and bulge); during the second episode, the thin disk formed 
almost independently of the previous infall episode, mainly out of material of 
primordial chemical composition. The disk formation process is assumed to be 
slower with increasing Galactocentric distance (Larson 1976; Matteucci \& 
Fran\c cois 1989); this ``inside-out'' formation of the Galactic disk is 
required to reproduce the abundance gradients and the gas distribution along 
the disk. A hiatus in the star formation between thick disk and thin disk 
phases, as suggested by recent observations (Gratton et al. 1996; Bernkopf 
\& Fuhrmann 1998; Fuhrmann 1998; Carraro 2000), is naturally produced by the 
model under the assumption that the star formation stops when the gas surface 
density drops below a threshold of 7 $M_\odot$ pc$^{-2}$ (Kennicutt 1989). 
Moreover, the presence of such a threshold predicts a roughly constant gas 
density profile at the outer radii (between 8 and 14 kpc) in agreement with 
observations (Dame 1993). We will focus on the thin disk formation process.
\par
The most relevant quantities in the present discussion are the infall rate and 
the SFR, since they involve the total mass surface density profile. The rate 
at which the thin disk is formed out of external matter (although it has also 
some initial contribution from the halo gas) is:
\begin{equation}
\frac{d\Sigma_I(R, t)}{dt} = B(R)\,e^{- (t - t_{max})/\tau_{D}},
\end{equation}
where $\Sigma_I(R, t)$ is the gas surface density of the infalling material, 
which has a primordial chemical composition; $t_{max}$ is the time of maximum 
gas accretion onto the disk, coincident with the end of the halo-thick disk 
phase which is set equal to 1 Gyr; $\tau_{D}$ is the timescale for mass 
accretion onto the thin disk component. We assume that $\tau_{D}$ increases 
with increasing the Galactic radius:
\begin{equation}
\tau_{D}(R) = 1.033 \times R - 1.267.
\end{equation}
This relation derives from the fact that in order to fit the G-dwarf 
metallicity distribution in the solar neighborhood it is necessary to assume 
$\tau_D$ = 7 Gyr, whereas to reproduce the metallicity distribution of the 
stars in the Galactic bulge ($R$ = 2 kpc) a timescale $\tau_D$ = 0.8 Gyr is 
required (Matteucci, Romano, \& Molaro 1999). It should be emphasized that 
this choice also guarantees a good fit to the gas density profile and to the 
Galactic abundance gradients. The quantity $B(R)$ is derived from the 
condition of reproducing the current total mass surface density distribution 
along the disk:
\begin{equation}
B(R) = \frac{\Sigma(R, t_{Gal})}{\tau_{D}(R) \Big( 1 - e^{-(t_{Gal} - 
t_{max})/\tau_{D}(R)} \Big)}.
\end{equation}
$t_{Gal}$ = 15 Gyr is the age of the Galaxy starting with the formation of 
the halo. From equation (3) it is clear the strong dependence of the infall 
rate from the total mass density profile.
\begin{table*}
\centering
{\footnotesize
\caption{Comparison between model predictions and observations for some
         relevant current quantities.}}
\vspace{0.5cm}
\small
\label{}
\begin{tabular}{c c c c c c c}
\tableline
\tableline
Quantity & 35 & 54 & 80 & 100 & Observations & Reference\\
\tableline
Type I SNe (century$^{-1})$&$0.27$&$0.37$&$0.59$&$0.72$&$0.17$--$0.7$&
$1$\\
Type II SNe (century$^{-1})$&$0.73$&$0.95$&$1.67$&$2.06$&$0.55$--$2.2$&
$1$\\
Nova outbursts (yr$^{-1}$)&$20$&$27$&$35$&$41$&$20$--$30$&
$2$\\
SFR\,($R_\odot, t_{Gal}$) ($M_\odot$ pc$^{-2}$ Gyr$^{-1}$)&$0$--$3$
&$2.57$&$2.89$&$3.58$&$2$--$10$&$3$\\
$\Sigma_{gas}\,(R_\odot, t_{Gal}$) ($M_\odot$ pc$^{-2}$)&$7.0$&$7.0$
&$8.4$&$10.3$&$7$--$13$&$4$\\
$\Sigma_{stars}\,(R_\odot, t_{Gal}$) ($M_\odot$ pc$^{-2}$)&$25.5$
&$38.4$&$54.3$&$65.6$&$35 \pm 5$&$5$\\
$\Sigma_{remnants}\,(R_\odot, t_{Gal}$) ($M_\odot$ pc$^{-2}$)&$2.5$
&$8.6$&$17.3$&$24.0$&$2$--$4$&$6$\\
$\dot{\Sigma}_{infall}\,(R_\odot, t_{Gal}$) ($M_\odot$ pc$^{-2}$ 
Gyr$^{-1}$)&$0.57$&$0.89$&$1.32$&$1.66$&$0.3$--$1.5$&$7$\\
\tableline
\end{tabular}
\small
\tablerefs{(1) Chiappini et al. 1997 and references therein; (2) Shafter 1997; 
	   (3) G\"usten \& Mezger 1982; (4) Dickey 1993; Flynn et al. 1999; 
	   (5) Gilmore et al. 1989; (6) M\'era, Chabrier, \& Schaeffer 1998 
	   quote $\Sigma_{WDs + NSs}$ = 2\,--\,4 $M_\odot$ pc$^{-2}$; here 
	   $\Sigma_{remnants}$ = $\Sigma_{WDs + NSs + BHs}$; (7) Portinari, 
	   Chiosi, \& Bressan 1998 and references therein}
\end{table*}
\par
Observations in our own and in other disk galaxies allow us to relate the SFR 
to intrinsic parameters of galaxies, in particular to the average surface gas 
density: SFR $\propto \Sigma_{gas}^k$ (Schmidt 1959), where the exponent is 
reasonably well known (e.g. $k$ = 1.4 $\pm$ 0.15 --- Kennicutt 1998). 
Moreover, the correlation between metallicity and surface brightness noted for 
late type spirals (McCall 1982; Edmunds \& Pagel 1984; Dopita \& Ryder 1994) 
is consistent with theories of self-regulated star formation, in which the 
energy produced by young stars and SNe feeds back into the ISM and inhibits 
further star formation by producing an expansion of the region surrounding 
the new stars or the SN remnants, thus relating the SFR to the gravitational 
potential and therefore to the total mass surface density.  In addition, 
according to the most popular prescriptions, one of the key conditions for 
star formation to occur is
\begin{equation}
t_{cooling} << t_{ff},
\end{equation}
where $t_{cooling}$ is the cooling timescale and $t_{ff}$ is the free-fall 
timescale (Katz 1992; Navarro \& White 1993; Carraro, Lia, \& Chiosi 1998); 
$t_{ff} \propto \rho^{-1/2}$, where $\rho$ is the total mass density (Buonomo 
et al. 1999). Again, in some way the SFR is related to the potential well and 
therefore to the total mass surface density. The SFR adopted here is the same 
as in Chiappini et al.\,(1997) and it was chosen in order to give the best 
agreement with the observed constraints:
\begin{displaymath}
\psi(R, t) = \nu(t)\,\left( \frac{\Sigma(R, t)}{\Sigma(R_\odot, t)} 
\right) ^{2\,(k - 1)}\,\left( \frac{\Sigma(R, t_{Gal})}{\Sigma(R, t)} \right)
^{k - 1} \cdot
\end{displaymath}
\begin{equation}
\cdot \Sigma_{gas}^{k}(R, t).
\end{equation}
$\nu(t)$ is the efficiency of the star formation process, which is set to 1
Gyr$^{-1}$, except when the gas surface density drops below 7 $M_\odot$ 
pc$^{-2}$. In fact in this case $\nu(t)$ = 0, because below this critical 
density the gas is gravitationally stable against density condensations into 
stars. $\Sigma(R, t)$ is the total disk mass surface density at a given radius 
$R$ and a given time $t$, $\Sigma(R_\odot, t)$ is the total disk mass surface 
density at the Solar position and $\Sigma_{gas}(R, t)$ is the gas surface 
density. Note that the gas surface density exponent, $k$, equal to 1.5, was 
obtained from the best model of Chiappini et al.\,(1997) in order to ensure 
a good fit to the observational constraints at the solar vicinity. This value 
is in good agreement with the recent observational results of Kennicutt (1998) 
and with N-body simulation results by Gerritsen \& Icke (1997).
\par
As far as our present model is concerned, we stress that the SFR \emph{along 
the Galactic disk}, and hence the evolutionary history of the Galactic disk 
itself, is fixed by the \emph{specific value of the local column density} of 
disk matter that we adopt. In fact, the total disk mass surface density 
profile, an exponential with scale length $R_{D}$, can be expressed as:
\begin{equation}
\Sigma(R, t_{Gal}) = \Sigma(R_\odot, t_{Gal}) \cdot e^{-(R - R_\odot)/R_{D}},
\end{equation}
which relates the column density at a radius $R$ to the column density at the 
Solar position. For the purpose of this paper, we let the local mass surface 
density value to vary from 35 to 100 $M_\odot$ pc$^{-2}$ and we set $R_\odot$ 
= 8 kpc.
\begin{table*}
\centering
{\footnotesize
\caption{ Theoretical solar abundances (mass fraction) are compared to
          the observed ones (Anders \& Grevesse 1989).}}
\vspace{0.5cm}
\small
\label{}
\begin{tabular}{c c c c}
\tableline
\tableline
X$_{\mathrm{i}}$ & This paper & Chiappini et al. 1997 & Observed\\
\tableline
H&$0.710$&$0.731$&$0.706$\\
D&$3.231 \times 10^{-5}$&$4.630 \times 10^{-5}$&$4.80 \times 10^{-5}$\\
$^3$He&$3.277 \times 10^{-5}$&$10.01 \times 10^{-5}$&$2.93 \times 10^{-5}$\\
$^4$He&$0.273$&$0.255$&$0.275$\\
$^7$Li&$8.740 \times 10^{-9}$&$...$&$9.63 \times 10^{-9}$\\
$^{12}$C&$3.920 \times 10^{-3}$&$1.827 \times 10^{-3}$&$3.03 \times 10^{-3}$\\
$^{13}$C&$5.215 \times 10^{-5}$&$4.758 \times 10^{-5}$&$3.65 \times 10^{-5}$\\
$^{14}$N&$1.670 \times 10^{-3}$&$1.386 \times 10^{-3}$&$1.11 \times 10^{-3}$\\
$^{16}$O&$7.378 \times 10^{-3}$&$7.278 \times 10^{-3}$&$9.59 \times 10^{-3}$\\
$^{20}$Ne&$9.729 \times 10^{-4}$&$9.42 \times 10^{-4}$&$1.62 \times 10^{-3}$\\
$^{24}$Mg&$2.563 \times 10^{-4}$&$2.48 \times 10^{-4}$&$5.15 \times 10^{-4}$\\
Si&$7.348 \times 10^{-4}$&$7.03 \times 10^{-4}$&$7.11 \times 10^{-4}$\\
S&$3.220 \times 10^{-4}$&$3.071 \times 10^{-4}$&$4.18 \times 10^{-4}$\\
Ca&$4.145 \times 10^{-5}$&$3.95 \times 10^{-5}$&$6.20 \times 10^{-5}$\\
Fe&$1.467 \times 10^{-3}$&$1.37 \times 10^{-3}$&$1.27 \times 10^{-3}$\\
Cu&$8.920 \times 10^{-7}$&$8.18 \times 10^{-7}$&$8.40 \times 10^{-7}$\\
Zn&$2.635 \times 10^{-6}$&$2.44 \times 10^{-6}$&$2.09 \times 10^{-6}$\\
Z&$1.700 \times 10^{-2}$&$1.433 \times 10^{-2}$&$1.886 \times 10^{-2}$\\
\tableline
\end{tabular}
\end{table*}
\par
Finally, the stellar nucleosynthesis prescriptions we adopt are from:
\begin{itemize}
\item[-] van den Hoek \& Groenewegen (1997) for low-intermediate mass stars 
         (0.8\,--\,8 $M_\odot$) (their case with variable $\eta_{AGB}$);
\item[-] Charbonnel \& do Nascimento (1998) for $^3$He production/destruction 
	 in low-mass stars ($M$ $<$ 2 $M_\odot$);
\item[-] Woosley \& Weaver (1995) for Type II SNe ($M$ $>$ 10 $M_\odot$);
\item[-] Thielemann, Nomoto, \& Hashimoto (1993) for Type Ia SNe (exploding 
	 white dwarfs in binary systems as described in Matteucci \& Greggio 
	 1986);
\item[-] Jos\'e \& Hernanz (1998) for classical novae.
\end{itemize}
Our prescriptions differ from those of Chiappini et al.\,(1997) in: i) the 
range of low-intermediate masses (0.8\,--\,8 $M_\odot$), where they adopted 
the Renzini \& Voli (1981) yields; ii) the fact that we are including the 
explosive nucleosynthesis from nova outbursts (Romano et al. 1999; Romano \& 
Matteucci 2000); iii) using extra-mixing prescriptions for the $^3$He 
nucleosynthesis (Chiappini \& Matteucci 2000) and iv) the Sun position, now 
taken as 8 kpc instead of 10 kpc.
\par
As far as the composition of the infalling material is concerned, we set 
Y$_{\mathrm{P}}$ = 0.245 rather than 0.23 (where Y$_{\mathrm{P}}$ is the 
primordial mass fraction in form of $^4$He), according to a recent estimate 
of the pristine $^4$He abundance obtained as a weighted mean of the $^4$He 
mass fractions from the two most metal-deficient blue compact galaxies, 
corrected for the stellar contribution (Y$_{\mathrm{P}}$ = 0.245 $\pm$ 0.002 
--- Izotov et al. 1999\footnote{Ionization corrections for low-metallicity 
H {\scriptsize II} regions suggest that this estimate should be reduced to 
Y$_{\mathrm{P}}$ = 0.241 $\pm$ 0.002 (Viegas, Gruenwald, \& Steigman 1999).}). 
The primordial abundances by mass of deuterium and helium-3 are 
D$_{\mathrm{P}}$ = 4.5 $\times$ 10$^{-5}$ and $^3$He$_{\mathrm{P}}$ = 4.0 
$\times$ 10$^{-5}$, respectively. The adopted IMF is that of Scalo (1986).

\section{MODEL PREDICTIONS}

\subsection{The Solar Circle}

In Tables 1 and 2 the model predictions together with the corresponding 
observed quantities are shown. Different models have identical parameters 
apart from the value of $\Sigma (R_\odot)$.
\par
From an inspection of Table 1, we see that there are not big differences among 
the quantities predicted by the different models, with the exception of the 
surface densities of live stars and stellar remnants, which seem to exclude 
the models with the highest total disk mass surface densities (80\,--\,100 
$M_\odot$ pc$^{-2}$). The lowest density value (35 $M_\odot$ pc$^{-2}$) 
predicts a stellar density which is only in marginal agreement with the 
observed one.
\par
The predictions on the abundances of the interstellar medium at the time of 
the Sun's formation are independent of the chosen total disk mass surface 
density but depend on all other model assumptions, in particular on the 
assumed stellar yields. Therefore, since we adopted a set of new yields 
relative to Chiappini et al.\,(1997), a comparison between our results for 
$\Sigma$ = 54 $M_\odot$ pc$^{-2}$ and those of Chiappini et al.\,(1997) is 
required. Table 2 shows that the largest differences in the predicted solar 
abundances arise for D, $^3$He, $^{12}$C, and $^{13}$C. The improvement in 
the predicted solar abundance for $^{12}$C and $^{13}$C is due to the fact 
that we adopt the new nucleosynthetic yields from van den Hoek \& Groenewegen 
(1997) for low and intermediate mass stars, whereas the improvement in the 
predicted solar abundances for D and $^3$He is due to the fact that we take 
into account that 93\% of evolved stars undergo the extra-mixing on the RGB 
and thus destroy, at least partly, their $^3$He (Charbonnel \& do Nascimento 
1998).
\par
One of the most severe observational constraints to chemical evolution 
models is the relative number of stars formed at a given time, i.e. at a 
given metallicity. In Fig.1 we compare the predicted and observed G-dwarf 
metallicity distributions; in particular, we show the results of models in 
which $\Sigma(R_\odot, t_{Gal})$ is taken to be 20, 54 and 100 $M_\odot$ 
pc$^{-2}$, respectively. Models with $\Sigma$ = 35 and 80 $M_\odot$ pc$^{-2}$ 
have also been computed. They produce theoretical G-dwarf distributions very 
similar to those obtained with $\Sigma$ = 54 or 100 $M_\odot$ pc$^{-2}$, so 
we do not show them here. From an inspection of Fig.1 we see that the case 
$\Sigma$ = 20 $M_\odot$ pc$^{-2}$ is in trouble in reproducing the high 
metallicity tail of the distribution while overestimating the number of low 
metallicity stars, whereas all the cases with $\Sigma$ lying between 35 and 
100 $M_\odot$ pc$^{-2}$ lead to a better agreement with the observations. We 
performed a $\chi^2$ test and obtained the following results: $\chi^2$ = 0.11 
for the case $\Sigma$ = 20 $M_\odot$ pc$^{-2}$, $\chi^2$ = 0.04 for the cases 
$\Sigma$ = 54 and 100 $M_\odot$ pc$^{-2}$. We conclude that the observed 
differential metallicity distribution is not a useful tool to distinguish 
among realistic values for the total baryonic mass density in the local disk. 
This is not surprising since the G-dwarf metallicity distribution depends 
mostly on the assumed law for the disk formation (i.e. on the assumed infall 
timescale).
\par
Fig.2 illustrates the different temporal behaviour of the SFR under different 
choices of $\Sigma$. In the low density case a stochastic SFR (between 0 and 3 
$M_\odot$ pc$^{-2}$ Gyr$^{-1}$ at the present time) is expected during most of 
the evolution, due to the fact that the threshold in the gas density, below 
which the star formation stops, is easily attained in this case. On the other 
hand, in the case of a very high $\Sigma$ the star formation process is not 
able to consume enough gas to go below the threshold. This explains why the 
highest $\Sigma$ values produce the high stellar and remnant densities listed 
in Table 1. In a recent paper Rocha-Pinto et al.\,(2000) suggest that
\vbox{ \vskip 0.8truecm
\centerline{\epsfxsize=8.6truecm
\epsfbox{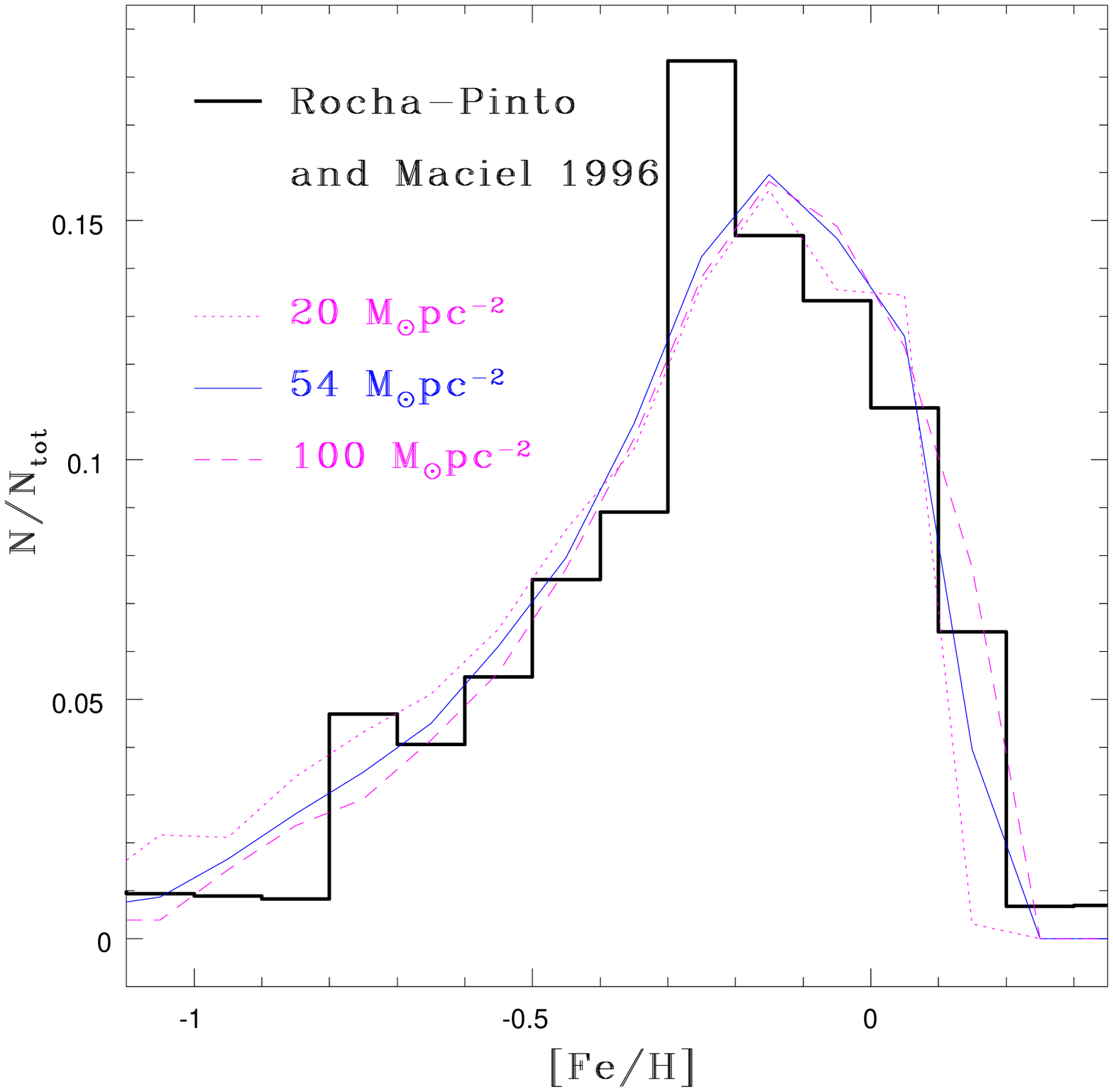}}
\vskip 0.00truecm
\figcaption[]{ Theoretical G-dwarf distributions obtained by 
               choosing different values of $\Sigma(R_\odot, 
               t_{Gal}$) as input parameters in the chemical 
               evolution model compared to the observed one 
	       (Rocha-Pinto \& Maciel 1996). The [Fe/H] ratios 
	       are normalised to the theoretical solar one for 
	       each model.}}
\vbox{
\centerline{\epsfxsize=8.6truecm
\epsfbox{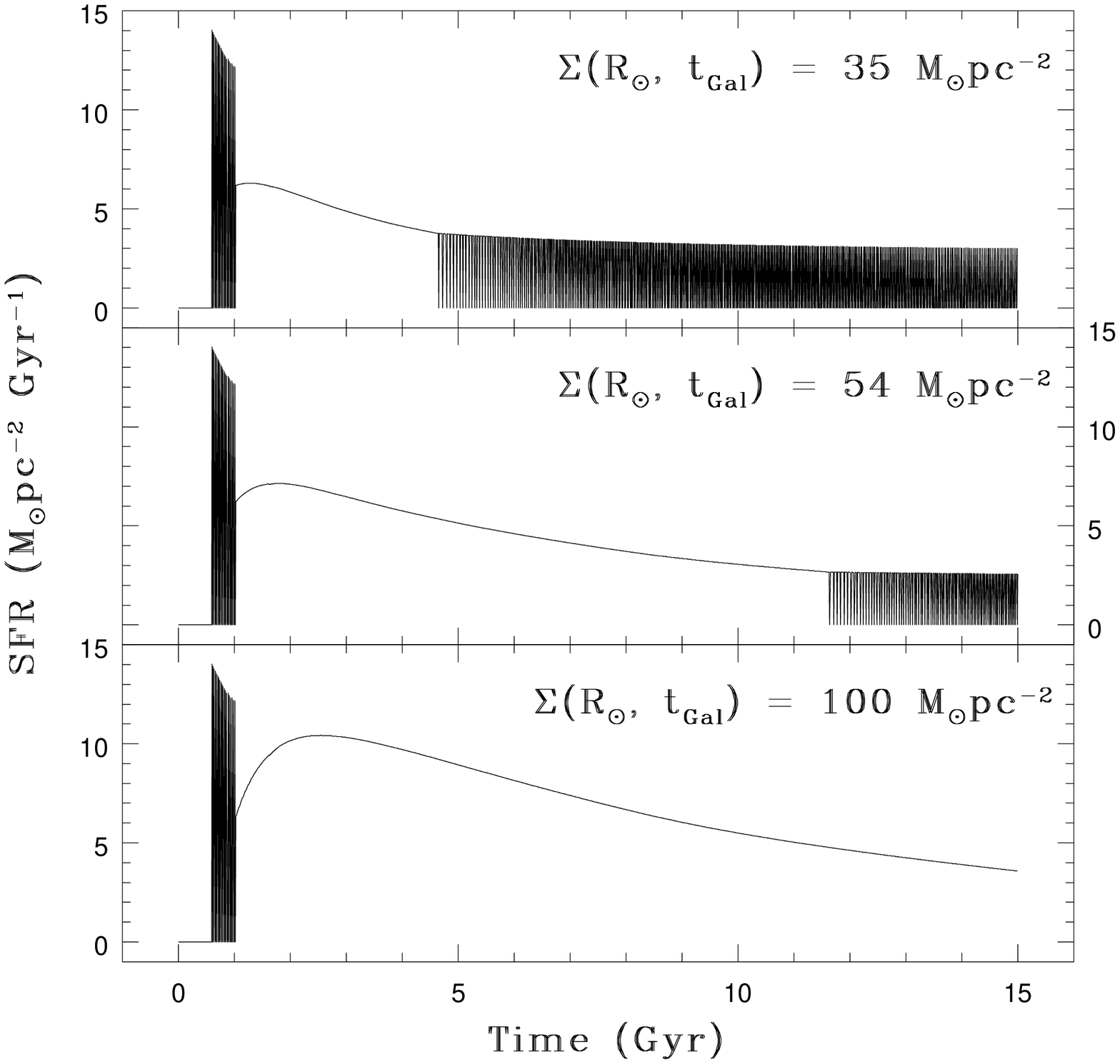}}
\vskip 0.00truecm
\figcaption[]{ Temporal evolution of the star formation rate 
               in the solar vicinity as predicted by models 
               adopting $\Sigma(R_\odot, t_{Gal})$ = 35, 54, 
               and 100 $M_\odot$ pc$^{-2}$.}
\vskip 0.8truecm }
the SFR in the disk should have 
proceeded in three distinct bursts at 0\,--\,1, 2\,--\,5, and 7\,--\,9 Gyr 
ago. The earliest episode of enhanced star formation should reflect the 
beginning of the solar vicinity formation (and probably corresponds to our 
maximum infall time). The other two episodes could be related to cyclical 
mechanisms (for instance, tidal interactions between the Galaxy and the 
Magellanic Clouds, or even induced star formation by spiral arms). As already 
explained, the theoretical G-dwarf metallicity distribution is mainly 
dependent on the assumed timescale for the formation of the solar vicinity, 
hence it would not be too much affected by recent bursts of star formation. 
\par
The elemental abundance ratios as a function of metallicity are nearly 
independent of the adopted $\Sigma$ value, depending mostly on the 
nucleosynthetic yields and the initial mass function. Therefore we do not 
discuss them here.

\subsection{Radial Profiles}

Here we discuss the predicted radial properties, in order to investigate the 
model behaviour along the overall Galactic disk under different prescriptions 
on the value of $\Sigma(R_\odot)$.
\par
We assume that the radial surface density distribution of the disk is 
exponential (see Sect.2, equation (6)), and run models assuming different 
values for the exponential scale length, $R_{D}$. Fig.3 shows the radial 
distribution of the present day surface gas density as predicted by models 
adopting $R_{D}$ = 2.5, 3.5, and 4.5 kpc (dashed, solid, and dotted lines, 
respectively), for different $\Sigma$ values. The model predictions are 
compared with the observed gas distribution derived by Dame (1993).
\vbox{ 
\centerline{\epsfxsize=8.6truecm
\epsfbox{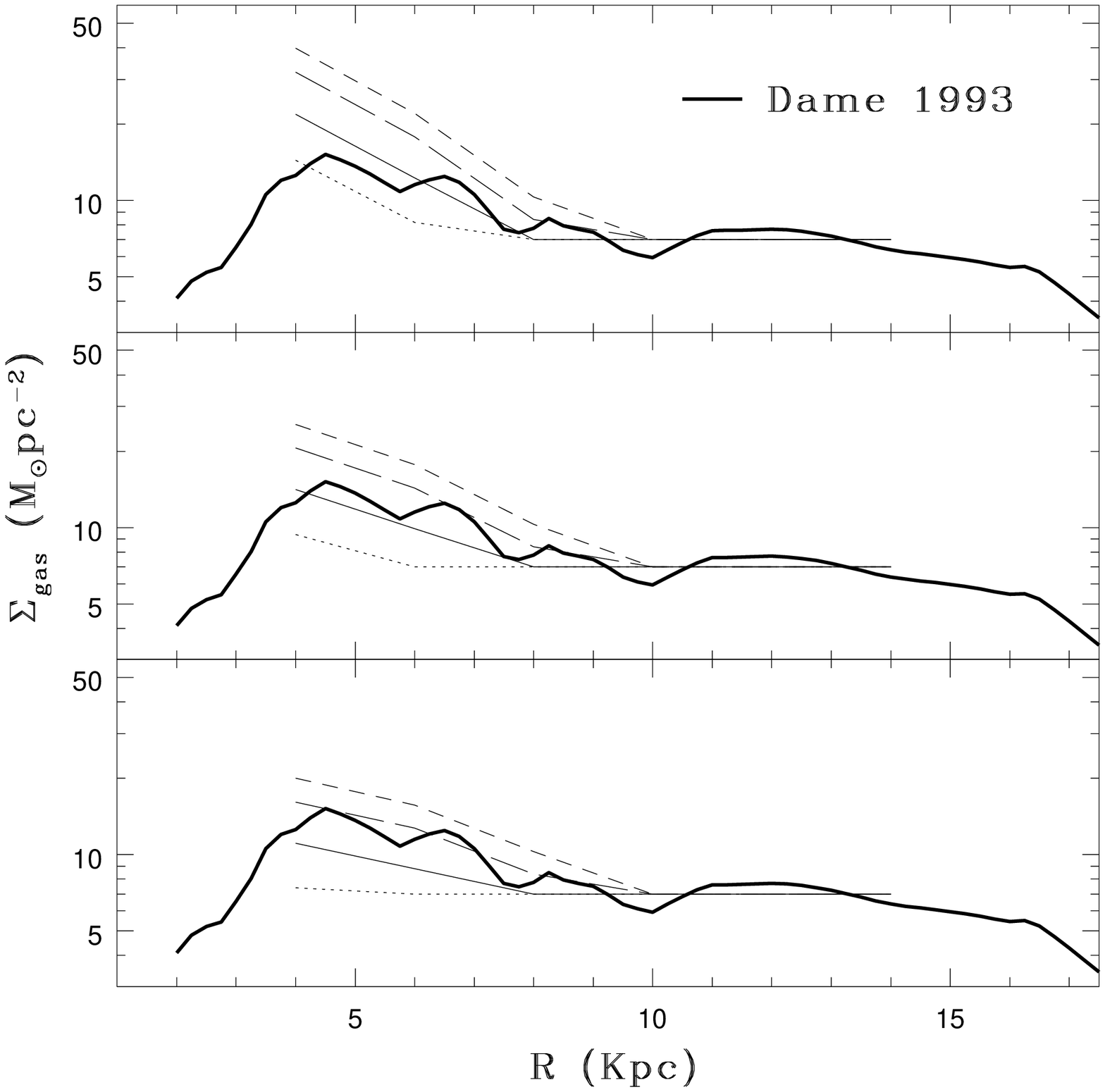}}
\vskip 0.00truecm
\figcaption[]{ Radial distribution of the present day 
	       surface gas density as predicted by models 
	       with R$_{D}$ = 2.5 (\emph{upper panel}), 
	       R$_{D}$ = 3.5 (\emph{middle panel}), and R$_{D}$ 
	       = 4.5 (\emph{lower panel}). From bottom 
	       to top, for each of these panels: lines referring 
	       to $\Sigma(R_\odot, t_{Gal})$ = 35, 54, 80, and
               100 $M_\odot$ pc$^{-2}$ (dotted, continuous, 
	       long-dashed, and short-dashed lines, respectively). 
	       The observed (H {\scriptsize I} + H$_2$) distribution 
	       is taken from Dame (1993). The surface density 
	       distribution of the total gas $\Sigma_{gas}$ 
	       is obtained from the sum of the H {\scriptsize I} 
	       and H$_2$ distributions, $\Sigma_{HI} + 
	       \Sigma_{H_2}$, accounting for the helium and 
	       heavy elements fractions (thick line).}
\vskip 0.8truecm }
\par
All our models predict a flattening of the distribution at radii larger than 
10 kpc, due to the presence of the threshold star formation density. Models 
with $\Sigma$ = 35 $M_\odot$ pc$^{-2}$ give a flat gas distribution already 
at the smallest radii; on the contrary, models with $\Sigma$ = 100 $M_\odot$ 
pc$^{-2}$ produce very steep distributions, hardly compatible with 
observations. All models predict that the gas content increases toward the 
Galactic center, whereas the observations clearly show a decreasing trend 
below 5 kpc. This problem could be solved by including the Galactic bar 
(e.g. Portinari \& Chiosi 2000), not present in the model. Taking into account 
these uncertainties we can conclude that the best gas distribution is obtained 
for local values of the total mass surface density in the range 50\,--\,75 
$M_\odot$ pc$^{-2}$.
\par
We also notice how increasing the disk scale length from 2.5 to 4.5 leads to 
a flattening in the predicted distribution. This behaviour is also observed 
in the case of the predicted radial density distributions of stars\footnote{ 
A scale length value of 3.5 kpc is the preferred one, being able to produce 
a final stellar density scale length in agreement with the observed one 
(2.5\,--\,3.0 kpc --- Sackett 1997; Freudenreich 1998).}, stellar remnants, 
and in the case of the predicted abundance gradients. In particular, higher 
total mass densities and larger scale lengths tend to flatten the gradients 
in the external regions of the disk. The actual situation is still 
controversial, with observations partly supporting this flattening 
(V\'\i lchez \& Esteban 1996; Maciel \& Quireza 1999) and partly not 
(Afflerbach, Churchwell, \& Werner 1997; Smartt \& Rolleston 1997; 
Rudolph et al. 1997). Therefore we can not draw any firm conclusion 
on this point.

\section{CONCLUSIONS}

We used a detailed model of galactic chemical evolution applied to our own 
galaxy in order to put constraints on the total amount of disk baryonic 
matter. Observational estimates of this quantity in the local disk span a 
wide range of values (from 35 to 100 $M_\odot$ pc$^{-2}$), depending on the 
underlying theoretical assumptions and on the adopted methods. We show that 
by means of chemical evolution models it is possible to substantially restrict 
the observed range of the total mass surface density.
\par
We show that the value of $\Sigma(R_\odot)$ should be restricted to the range 
50\,--\,75 $M_\odot$ pc$^{-2}$, in order to guarantee the best fit to all the 
observational constraints available for the solar neighborhood and the overall 
Galactic disk. This is well consistent with the mass range which Cr\'ez\'e et 
al.\,(1998) and Holmberg \& Flynn (1998) advocate from their analysis of A and 
F stars with parallaxes and proper motions from the Hipparcos satellite.

\acknowledgements
We would like to thank Dennis W. Sciama for enlightening comments and Cedric 
G. Lacey for careful reading of the manuscript. We would like also to thank 
the referee, Dr. Chris Flynn, for many comments that improved the presentation 
of the paper. D.R. wish to thank Cesario Lia for discussing the feed-back 
mechanism. F.M. and P.S. acknowledge financial support from the Italian 
Ministry for University and for Scientific and Technological Research (MURST). 
C.C. wish to thank Thomas M. Dame for having kindly sent his data on the gas 
density ditribution along the disk and to acknowledge financial support from 
CNPq/Brazil.

\vfill\eject

\end{document}